\documentclass[11pt,a4paper]{article}
\pdfoutput=1
\usepackage{jcappub}
\usepackage{graphicx,psfrag,color}
\usepackage{bm}
\usepackage{slashed}
\usepackage{amsmath,amssymb,mathtools}
\usepackage{multirow}
\usepackage{color,ulem}
\usepackage{graphicx}
\allowdisplaybreaks
\usepackage[colorlinks=true,linkcolor=blue,citecolor=blue]{hyperref}%

\begin{document}

\begin{flushright}
ULB-TH/16-16
\end{flushright}

\title{Supernova Constraints on Massive (Pseudo)Scalar Coupling to Neutrinos}


\author[a]{Lucien Heurtier,}
\author[a,b]{Yongchao Zhang}
\affiliation[a]{Service de Physique Th\'{e}orique, Universit\'{e} Libre de Bruxelles, Boulevard du Triomphe, CP225, 1050 Brussels, Belgium}
\affiliation[b]{School of Physics, Sun Yat-Sen University, Guangzhou 510275, China}

\emailAdd{yongchao.zhang@ulb.ac.be}
\emailAdd{lucien.heurtier@ulb.ac.be}

\date{\today}

\abstract{
  In this paper we derive constraints on the emission of a massive (pseudo)scalar {$S$} from annihilation of  neutrinos in the core of supernovae through the dimension-4 coupling $\nu\nu S$, as well as the effective dimension-5 operator $\frac{1}{\Lambda}(\nu\nu)(SS)$. While most of earlier studies have focused on massless {or ultralight} scalars, our analysis involves scalar with masses of order $\mathrm{eV- GeV}$ which can be copiously produced during {the explosion of supernovae, whose core temperature is} generally of order $T\sim \mathcal{O}(10)$ MeV. From the luminosity and deleptonization arguments
  regarding the observation of SN1987A, we exclude a large range of couplings $ 10^{-12} \lesssim {|g_{\alpha\beta}|}\lesssim  10^{-5}$ for the dimension-4 case, depending on the neutrino flavours involved and the scalar mass. In the case of dimension-5 operator, for a scalar mass from MeV to 100 MeV the coupling $h_{\alpha\beta}$ get constrained from $10^{-6}$ to $10^{-2}$, with the cutoff scale explicitly set $\Lambda = 1$ TeV. We finally show that if the neutrino burst  {of} a nearby supernova explosion {is} detected by Super-Kamiokande and  IceCube, the constraints {will} be largely reinforced.
}
\maketitle


\section{Introduction}

Among other problems, the standard model (SM) of particle physics suffers from lots of mysteries concerning the neutrino sector.  As a matter of fact, the three neutrino flavours have been established to oscillate as they propagate in vacuum or matter and hence be split in mass. A large variety of current experiments aim to detect neutrinos produced in the Sun, atmosphere, reactors, nuclei decay or astrophysical sources. The next generation neutrino experiments, as well as collider studies, may help us pin down some of the fundamental mass and mixing parameters, resolve the mystery of Majorana versus Dirac nature of neutrinos, and reveal hints about the neutrino mass generation mechanisms.

As a peculiar source of astrophysical neutrinos, supernovae provide an alternative environment to study neutrinos and their interplay to nuclear physics and possibly new, feeble, secret interactions. Indeed, such interactions could drastically alter the evolution of astrophysical objects, and in particular the dynamics of supernova explosions, and be possibly measured by neutrino detectors on earth. There has been an extensive study regarding the SM together with hypothetical beyond SM particles in the evolution of supernovae, including {both} active and sterile neutrinos~\cite{Shi:1993ee,Pastor:1994nx,Kolb:1996pa,Nunokawa:1997ct,Caldwell:1999zk,
Fetter:2002xx,Keranen:2004rg,Beun:2006ka,Choubey:2007ga,Hidaka:2007se,Keranen:2007ga,
Fuller:2009zz,Raffelt:2011nc,Tamborra:2011is,Wu:2013gxa,Esmaili:2014gya,Warren:2014qza,
Zhou:2015jha,Arguelles:2016uwb,Minakata:1989mk,Hidaka:2006sg,Warren:2016slz,Mathews:2016xba}, (very) light pseudoscalars (e.g. axion~\cite{Turner:1987by,Brinkmann:1988vi,
Mayle:1987as,Burrows:1988ah,Mayle:1989yx,
Turner:1988bt,Kolb:1988pe,Raffelt:1991pw,Janka:1995ir,Keil:1996ju,Raffelt:2006cw,
Giannotti:2010ty,Fischer:2016cyd} and Majoron~\cite{Gelmini:1982rr,Kolb:1987qy,
Choi:1987sd,Berezhiani:1989za,Choi:1989hi,Chang:1993yp,Kachelriess:2000qc,
Tomas:2001dh,Lindner:2001th,Hannestad:2002ff,Farzan:2002wx,Fogli:2004gy,
Das:2011yh}) as well as vector bosons (dark photon~\cite{Kolb:1996pa,Dent:2012mx,Dreiner:2013mua,Kazanas:2014mca}), or even dark matter (DM) particles~\cite{Minakata:1989mk,Hidaka:2006sg,Warren:2016slz,Mathews:2016xba,
Fayet:2006sa,Zhang:2014wra,Blackadder:2014wpa,Graham:2015apa,Bramante:2015cua,
Guha:2015kka,Brdar:2016ifs}.

In this paper we aim to focus on the supernova constraints in the neutrino sector, in the presence of a massive scalar $S$ or pseudoscalar $J$ coupled to the SM neutrinos. A well motivated example of such {particles} is the Majoron~\cite{Chikashige:1980qk,Chikashige:1980ui,Aulakh:1982yn,Gelmini:1980re,Schechter:1981cv}, the Goldstone boson generated from global lepton number symmetry breakdown. This particle is intimately related to neutrino mass generation via the so called seesaw mechanisms~\cite{type1a,type1b,type1c,type1d,type1e,
type2a,type2b,type2c,type2d,type3} and the lepton number breaking scale. However, the Majoron does not have to be exactly massless~\cite{Akhmedov:1992hi,Rothstein:1992rh} nor a Goldstone boson~\cite{Bamert:1994hb} as in the original models, and could even play the role of DM~\cite{Berezinsky:1993fm,Starkman:1993ik,Dolgov:1995rm,Arai:1998ni,
SommerLarsen:1999jx,Kazanas:2004kv,Lattanzi:2007ux,Lattanzi:2008ds,Bazzocchi:2008fh,
Aranda:2009yb,Gu:2010ys,Esteves:2010sh,Ghosh:2011qc,Lattanzi:2013uza,Queiroz:2014yna,
Boucenna:2014uma,Lattanzi:2014mia,Dudas:2014bca,Dutra:2015vca,Escudero:2016tzx} or dark radiation~\cite{Chang:2013yva,Chang:2014lxa,Chang:2016pya}. Such a light (pseudo)scalar interacting with neutrinos could also emerge in a large range of beyond SM scenarios, such as supersymmetric and extra dimension models~\cite{Bamert:1994hb,Mohapatra:1988fk,Burgess:1992dt,Montero:2000ar,
Mohapatra:2000px}, and could also have large couplings to neutrinos in, for instance, modified seesaw models involving large flavour violation \cite{Heurtier:2016iac}.

Without intending to dip into any of these specific phenomenological models but rather investigating in a model independent manner how supernovae can constrain the couplings of light scalar bosons to the active neutrinos~\cite{Pasquini:2015fjv}, the dimension-4 and dimension-5 Lagrangians given respectively in Eqs.~(\ref{eqn:coupling4}) and (\ref{eqn:coupling5}) can be viewed as the low energy remnant of UV complete underlying theories. Setups involving peculiar global symmetries under which both leptons and the Majoron field are charged (such as the original Majoron model in  type I seesaw) can naturally provide couplings of the (pseudo)scalar to neutrinos at the tree level whereas the coupling to charged leptons arise at the loop level. An example of such model can be found for instance in~\cite{Heurtier:2016iac} where the possibility of a massive scalar coupling to neutrinos is in tension with the prospective bounds we  present in this paper. 
Until now the literature has mostly focused on massless or very light ($\lesssim\mathrm{eV}$) particles, in the quest for Majorons or Axion-Like Particles, we will here release a complementary study, exploring masses of the eV-GeV range, since as we will see, such a broad region is to be particularly constrained.

In some of the models involving light scalars in the neutrino sector, there exist also the heavy right-handed neutrinos (RHNs), which are used to generate the tiny active neutrino masses via the seesaw mechanism.
One should however comment on the fact that in a certain class of seesaw models, our study cannot be applied. For instance in the case of a standard type I seesaw, writing the lagrangian as follows
\begin{equation}
\mathcal{L}\supset -\lambda \phi \bar N_R^c N_R - y\bar L H N_R -V(H,\,\phi)\,,
\end{equation}
where $H$ is the SM doublet and $\phi=v_S+(S+iJ)/\sqrt 2$, the masses of the scalar and RHNs are given respectively by
\begin{equation}
m_S=\sqrt {\lambda_S} v_S\,\text{~~and~~} M_N = \lambda v_S \,,
\end{equation}
with $\lambda_S$ quartic coupling in the potential $V$. Thus for an $\mathcal O(1)$ parameter $\lambda$ and a small quartic coupling $\lambda_S$ one can easily be in a situation where the RHNs are far heavier than the scalar $S$. In such a situation, if the RHNs are heavier than the GeV scale (or even hundreds of MeV) it can not be produced in the core of the supernovae and hence will not be present in any of the calculations below. On the opposite, the cases containing low mass RHNs would imply that the presence of RHN's would affect the neutrinos thermal distribution as well as the decay length and mean free path of the scalar $S$. Such consideration would strongly rely on the choice of the seesaw model considered and thus be very model dependent and require a proper modelling of the RHNs distribution in the core. We will for now on consider this case as being out of the scope of our paper, and focus on cases where all RHN's are heavier than the GeV scale. One may anyway note that, RHNs participating to leptogenesis have generically to be heavier than 100 MeV not to hit BBN bounds \cite{Canetti:2012kh, Hambye:2016sby}, which make our study rather general unless one tries to explore the possibility of a RHN dark matter.

In the presence of massive degrees of freedom such as the scalar $S$, the evolution of supernova neutrino bursts and the subsequent deleptonization phase will be substantially affected, if the couplings to neutrinos are sufficiently large, and thus one can set limits on these couplings from the observation of SN1987A~\cite{Hirata:1987hu,Bionta:1987qt} as well as a future supernova explosion. Two different types of constraints can be obtained if one assumes that the extra scalar boson couples solely to the SM neutrinos: (i) On the one hand, the energy loss due to emission of (pseudo)scalars from the supernova core could significantly reduce the total flux of neutrinos observed in supernova explosions. For SN1987A the neutrino energy depletion is of order $10^{53}$ erg, and it is expected that the energy carried away by exotic particles could not exceed a sizable fraction of it. (ii) On the other hand, the deleptonization effect, e.g. $\nu_e \nu_e \to J$ can dramatically disable the explosion of supernovae. Indeed, supernova simulations reveal that the explosion process is very sensitive to the electron neutrino fraction $Y_{L_e}$ inside the core~\cite{Bruenn:1985,Baron:1987zz}.
Thus by evaluating the number emission rate of the light scalars, we can set limits on the couplings to neutrinos involving the electron flavor, which depends yet largely on the supernova modelling and simulation details.

The supernova constraints for a massless pseudoscalar Majoron have been extensively studied in Ref.~\cite{Gelmini:1982rr,Kolb:1987qy,
Choi:1987sd,Berezhiani:1989za,Choi:1989hi,Chang:1993yp,Kachelriess:2000qc,
Tomas:2001dh,Lindner:2001th,Hannestad:2002ff,Farzan:2002wx,Fogli:2004gy,
Das:2011yh} where the Majoron is produced via the lepton number violating processes $\nu + \nu \to J$ and $\nu \to \bar \nu + J$, rendered possible through matter effects. The limits come out to be $g_{\alpha\beta} \lesssim 10^{-5}$,
from both the luminosity and deleptonization arguments. In the case of a massive scalar, if the scalar mass is much larger than the matter effects, then the calculation procedure is somewhat similar to the case of heavy axion or dark photon, with the significant difference that the two latters are produced from nucleon collisions. On the other hand, for a supernova core with temperature $T \simeq 10-30$ MeV, the scalars can not be produced abundantly for masses above $\gtrsim$ GeV, which is heavily suppressed by the factor of $e^{- m / T}$. Thus the limits we will present in this paper apply to scalars in the mass range of eV $\lesssim m \lesssim$ GeV.

A massless Majoron could also be emitted from neutrinoless double beta decays~\cite{Mohapatra:1988fk,Chang:1993yp,Bamert:1994hb,Burgess:1992dt,
Montero:2000ar,Mohapatra:2000px,Berezhiani:1992cd,Burgess:1993xh,Hirsch:1995in} 
and from the decays of SM mesons and leptons~\cite{Gelmini:1982rr,Barger:1981vd,Britton:1993cj,Lessa:2007up,Pasquini:2015fjv}.
Ref.~\cite{Pasquini:2015fjv} {derived} recently the limits from meson and lepton decays {in} the case of a massive scalar, with mass up to $\sim 100$ MeV, excluding couplings of order $|g_e|^2\sim 10^{-1}-10^{-6}$. The constraints on massless and massive scalars are complementary to each other, and we will see that supernova constraints for the massive case in this paper will push the couplings down by several orders of magnitude.

For the sake of completeness we will consider, in addition to the usual dimension-4 couplings of the form $\nu \nu S$, the possibility of having non-renormalizable dimension-5 operators of the form $(\nu \nu) (S S) / \Lambda$, where $\Lambda$ stands for a cutoff scale after the heavy UV complete sector is integrated out, and in this case the scalar $S$ can be considered for instance to be a neutrino-phillic light DM candidate, whose couplings to other SM particles 
are vanishing or highly suppressed.
All other higher order effective operators are less important from the phenomenological point of view in this paper and will be neglected.

The paper is organized as follows: The next section is devoted to the production and decay of the massive scalar in supernova core induced by the dimension-4 couplings. The luminosity limits and trapping effect are presented respectively in Section~\ref{sec:lum} and \ref{sec:mfp}, while the deleptonization constraints are considered in Section~\ref{sec:delep}. The analogous constraints on dimension-5 interactions are derived and collected in Section~\ref{sec:dim5}.  {Finally, future prospects for possible   Super-Kamiokande and IceCube detections are given in Section~\ref{sec:prospects}}, before we summarize and conclude in Section~\ref{sec:conclusion}. Some of the calculation details are listed in the Appendix.


\section{Scalar production and decay}
\label{sec:production}

For a {massive} scalar $S$ or pseudoscalar $J$, the {renormalizable} couplings to neutrinos can be generally denoted by
\begin{eqnarray}
\label{eqn:coupling4}
\mathcal{L} =
\frac12  g_{\alpha\beta} S \left( \nu^T_\alpha i\sigma_2 \nu_\beta \right) +
\frac12 g'_{\alpha\beta} J \left( \nu^T_\alpha \sigma_2 \nu_\beta \right)+ {\rm h.c.}
\end{eqnarray}
where $\alpha$ and $\beta$ are flavour indices. The coupling matrices {$g^{(\prime)}_{\alpha\beta}$} are symmetric in the flavor basis and can be rotated into the mass basis via
\begin{eqnarray}
g^{(\prime)} \to U g^{(\prime)} U^T
\end{eqnarray}
with $U$ the PMNS matrix connecting the two sets of basis. It should be noted that the couplings in Eq.~(\ref{eqn:coupling4}) could be either lepton number conserving or violating by one or two, depending on the lepton numbers of scalars which could be either 0, 1 or 2~\cite{Bamert:1994hb,Hirsch:1995in}. The lepton number assignments are, however, rather  model dependent, and we will not go into such details. Moreover, in the calculation of the energy and number emission rates below the scalars and pseudoscalar interactions do not show any difference, thus we will proceed from now on with the scalar $S$ without loss of generality.


As far as the dimension-4 couplings are concerned, the dominant production {mechanism} of $S$ comes from the {annihilation of neutrinos} $\nu_\alpha \nu_\beta \to S$. {The double production process} $\nu \bar\nu \to SS$ are suppressed by the higher  {powers} of the couplings $g^2_{\alpha\beta}$ and can be safely neglected throughout the paper.\footnote{Note that even in the case of a massless scalar, the inverse decay process is kinematically forbidden in the vacuum but can occur via matter effects, and turns out to dominate also over the $2\to2$ processes~\cite{Farzan:2002wx}.}
Let us assume for simplicity that the { scalar $S$} decays only into the SM neutrinos, with the partial width in the supernova core frame
\begin{eqnarray}
\Gamma_{\alpha\beta} &\equiv& 2\Gamma (S \to \nu_\alpha \nu_\beta)
= \frac{|g_{\alpha\beta}|^2 m_S}{8\pi} \frac{m_S}{E_S} \,. 
\end{eqnarray}
where the factor of 2 takes into account {the anti-neutrino channels}, and $E_S/ m_S$ is the time dilation factor.

\section{Luminosity constraints}
\label{sec:lum}

The density in  {the} supernova core is very high, at the level of $10^{14} \, {\rm gram}/{\rm cm}^3$, and a huge number of neutrinos are produced from the SM weak processes. It is expected that a large number of light scalars can be produced in the neutrino bath before the explosion is ignited and the emission of neutrino burst. If the couplings are small, the scalars will travel easily outside the inner core, and carry away a sizable portion of the binding energy. The total neutrino luminosity from SN1987A within one second after {explosion} is about $\mathcal{L}_\nu \simeq 5\times 10^{52}$ erg/s, which is consistent with the theoretical predictions of supernova models,
{thus the scalar mass and couplings to neutrinos are severely constrained from the simple luminosity argument.}


For the production process $\nu_\alpha (p_1) \nu_\beta (p_2) \rightarrow S(p_S)$, the energy emission rate per unit volume of supernova core is given by~\cite{Dent:2012mx,Kazanas:2014mca}:
\begin{eqnarray}
\label{eqn:production}
Q_{} = \int d\Pi_3 \ F_S \sum_{\rm spins}|\mathcal{M}|^2(2\pi)^4 \delta^{(4)}(p_1+p_2-p_S) E_S f_1 f_2 \,,
\end{eqnarray}
where $d \Pi_3$ is the phase space for the incoming and outgoing particles, $F_S = 1/(1+ \delta_{\alpha\beta})$ is the symmetry factor, $\cal M$ the production amplitude, and
\begin{eqnarray}
f_i (E_i) = \frac{1}{1 + e^{(E_i - \mu_i)/T}}
\end{eqnarray}
the initial-state Fermi-Dirac distribution with $\mu_i$ the chemical potential. After straightforward simplifications, the energy emission rate is
\begin{eqnarray}
\label{eqn:Q}
Q_{}
&=& \frac{F_S |g_{\alpha\beta}|^2 T^5}{2^5 \pi^3} \int_{q^2/16v}^{\infty} du \int_{0}^{\infty} dv \, \frac{(\sqrt{u}+\sqrt{v})q}{2\sqrt{uv}} \,
\frac{1}{(1 + e^{\sqrt{u} - x_1})(1 + e^{\sqrt{v} - x_2})} \,,
\end{eqnarray}
where we have defined the dimensionless variables $u \equiv {\bf p}_{\bf 1}^2/T^2$, $v \equiv {\bf p}_{\bf 2}^2/T^2$, $q \equiv {m_S^2}/{T^2}$ and $x_i \equiv {\mu_i}/{T}$ with the core temperature $T = 30$ MeV.
Once produced, the scalars can decay back into neutrinos. If the scalars decay inside the core, then no {exotic} energy depletion can occur, {and} the decay factor $e^{-\Gamma_{\alpha\beta} R_C}$ has to be plugged into Eq.~(\ref{eqn:Q}) with $R_C = 10$ km the core radius.
Multiplying further  the total core volume $V_{\rm core} = \frac43 \pi R_C^3$, we obtain the total luminosity due to the scalar emission.

\begin{figure*}[t]
  \centering
  \includegraphics[width=0.48\textwidth]{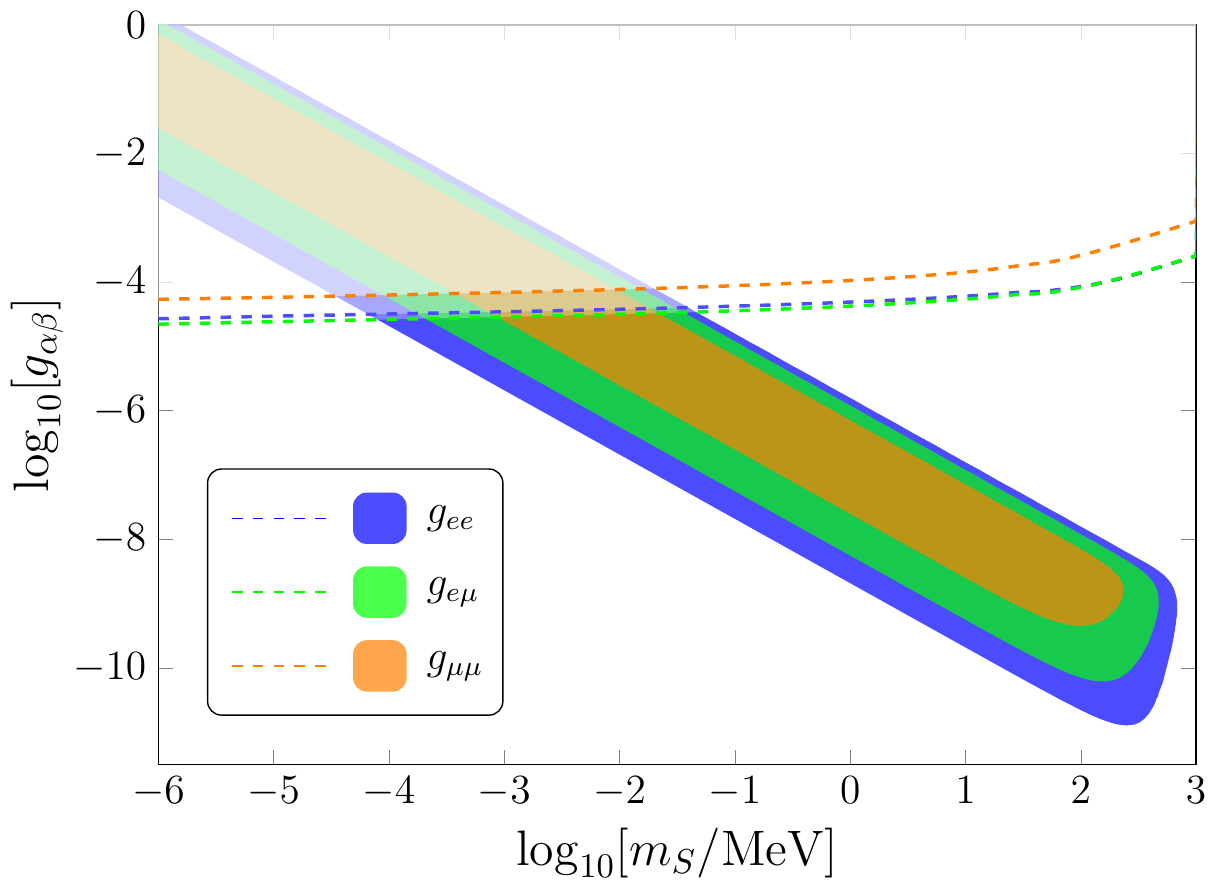}
  \includegraphics[width=0.48\textwidth]{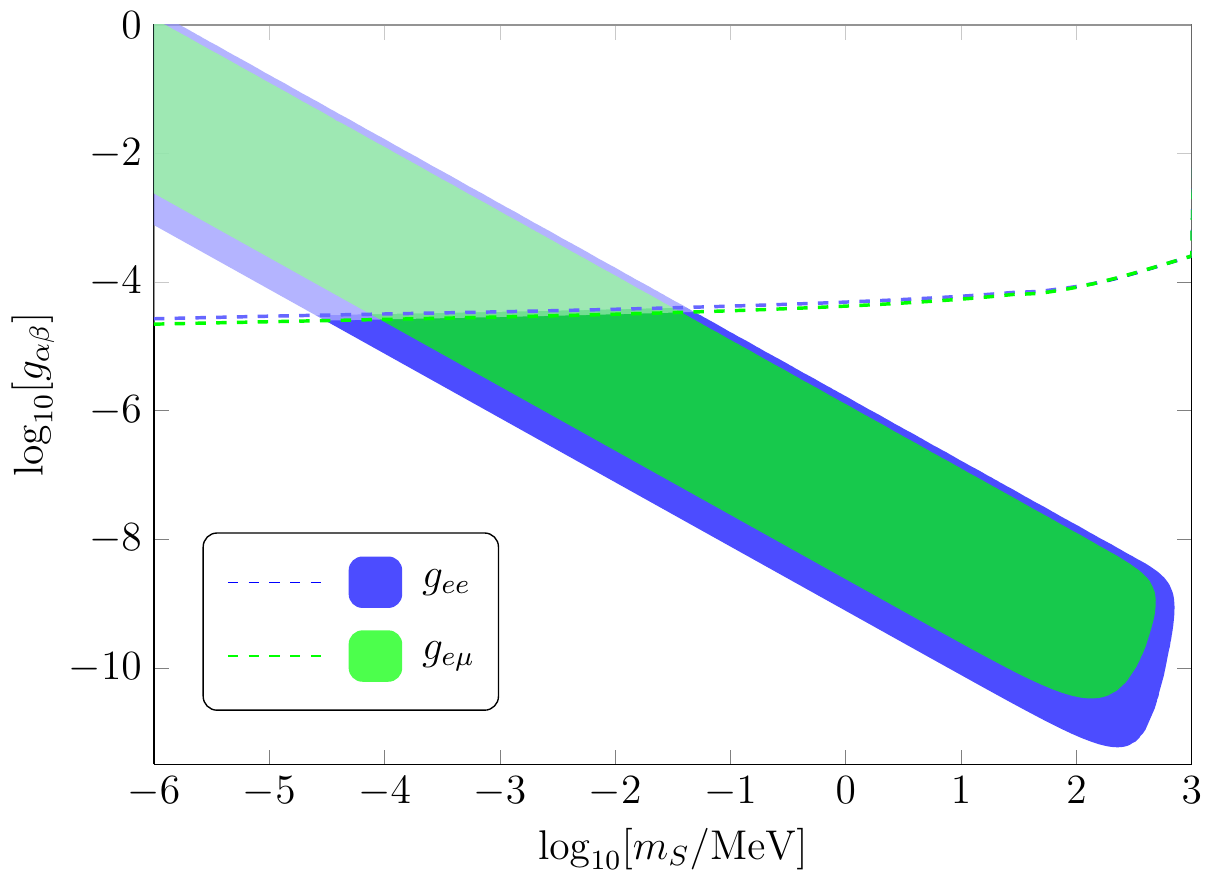}
  \caption{\label{fig:lum}\footnotesize{
  {\it Left panel:} Neutrino luminosity constraints on the coupling $g_{\alpha\beta}$ as functions of the scalar mass $m_S$. The coloured shaded region are excluded by the luminosity condition whereas in the lighter color regions above the dashed lines the scalar mean free path $D_S$ is smaller than the core radius $R_C$. {\it Right panel:} Constraints on the couplings $g_{e\beta}$ involving the electron flavour, using the deleptonization condition, as functions of the scalar mass $m_S$. The constraints involving the $\tau$ flavour is the same as that for the $\mu$ flavour.} }
\end{figure*}

All the constraints on $m_S$ and $|g_{\alpha\beta}|$ are presented in Fig.~\ref{fig:lum}, where the shaded regions are excluded. To make things as simple as possible, we switch on only one entry of $g_{\alpha\beta}$ and set all other independent elements to be zero. In the supernova {core}, a huge number of $\nu_e$ are emitted in the electron capture process, pushing the chemical potential {of} $\nu_e$ to be rather large, $\sim$200 MeV, whereas {for} $\bar\nu_e$ it is the negative values of $-200$ MeV. On the other hand,  in first approximation the $\mu$ and $\tau$ flavors are completely on the same footing, and thus we assume their chemical potentials to be zero. As demonstrated in \cite{Raffelt:1996wa}, the typical time scale of neutrino oscillations are constrained by experiments to be much longer than the typical time scale of the neutrino burst immediately after the core collapse. Therefore the number density hierarchy of different neutrino species remains relatively constant before the core explosion.
Thus the luminosity {$\mathcal{L}_S$ } of  {scalar} production in the supernova core is expected to have a hierarchical structure in the flavour basis:
\begin{eqnarray}
\mathcal{L}_S (\nu_e \nu_e \to S) \gg
\mathcal{L}_S (\nu_e \nu_{\mu,\tau} \to S) \gg
\mathcal{L}_S (\nu_{\mu,\tau} \nu_{\mu,\tau} \to S) \,. \nonumber
\end{eqnarray}
This hierarchy explains the repartition of constraints presented in Fig.~\ref{fig:lum}: the constraints involving $\nu_e$ are more stringent and exclude larger regions, whereas the constraints involving the $\mu$ and $\tau$ flavours are similar and somehow less stringent. As previously mentioned, when the scalar is very heavy, say $\sim$ GeV, it can not be produced abundantly in the supernova core, and thus can not deplete much of the energy. On the other hand, in Fig.~\ref{fig:lum} its mass is down to the eV scale, at which the matter effects in the highly dense core might become important and thus can not be naively neglected.

\section{Trapping effect}
\label{sec:mfp}

After production in the supernova core, we have seen that the scalars can decay back into neutrino. This effect has already been taken into account in the energy emission rate as described above. Another possible issue is that the scalar scatters on the neutrino bath inside the supernova core via $\nu_\alpha + S \to \nu_\beta + S$\footnote{{The process $\nu + S \to \bar\nu$ is kinematically forbidden in the approximation of vanishing matter effect and neutrino masses. Since we are considering scalar mass $\gtrsim$ eV, this assumption remains true in all our calculations.}}, mediated by a neutrino $\nu_\gamma$ in the $s$ and $u$ channels. It is straightforward to calculate the scattering cross section,
\begin{eqnarray}
\sigma = \frac{1}{64 \pi E_S (p_\nu + E_S)} \int d  \cos\theta
\left| \mathcal{M} \right|^2 \,.
\end{eqnarray}
Then the mean free path (MFP) for scalars {propagating in the supernova core} is given by
\begin{eqnarray}
D_S^{-1} = \frac{ \sum_{\alpha,\, \beta} \int d E_S \, n_{\nu_\alpha} \, \sigma (\nu_\alpha S \to \nu_\beta S ) f (E_S)}{\int d E_S \, f (E_S)} \,,
\end{eqnarray}
where in general both flavours in the initial and final states have to be summed up. For simplicity we assume the standard number density distribution for the incoming scattering {neutrinos}
\begin{eqnarray}
n_\nu = \frac{2}{(2\pi)^3} \int \frac{2\pi p^2 dp \, d \cos\theta}{e^{(E-\mu)/T} + 1}
\end{eqnarray}
with $\cos\theta$ the angle between the scattering neutrino and scalar impulsion in the rest frame of supernova core. At leading order the scalar is produced from annihilation of neutrinos, thus we can estimate the scalar energy distribution function $f(E_S)$ from convolution of two neutrino distributions $f(E_{\nu_\alpha})$ and $f(E_{\nu_\beta})$. It turns out that
\begin{eqnarray}
\label{eqn:fES}
f(E_S) = \frac{e^{(\mu_\alpha + \mu_\beta)/T}}{e^{(\mu_\alpha + \mu_\beta)/T} -e^{E_S/T}}
\log \left[ \frac{\left(1+e^{\mu_\alpha/T}\right) e^{\mu_\beta/T}}{e^{\mu_\beta/T} + e^{E_S/T}} \right] \,.
\end{eqnarray}
To calculate the MFP, we integrate over the momentum/energy of the incoming neutrinos and scalars:
\begin{eqnarray}
D_S^{-1} &=& \frac{\sum_{\alpha,\, \beta}
\int p_{\nu_\alpha}^2 d p_{\nu_\alpha} d E_S \, d \cos\theta \,
f (E_{\nu_\alpha}) f(E_S) \,
 \sigma (\nu_\alpha S \to \nu_\beta S )} {2\pi^2 \int d E_S \, f (E_S) } \,.
\end{eqnarray}

{All the MFP limits $D_S = R_C$ are  {indicated}   in Fig.~\ref{fig:lum} by dashed lines,  {above which the MFP is smaller than the core radius}. As for the luminosity argument, only one independent element of the coupling matrix $g_{\alpha\beta}$ is assumed to be non-vanishing at one time.}
Since the scattering $\nu S \to \nu S$ has a quartic dependence on the coupling $g_{\alpha\beta}$, the MFP condition $D_S < R_C$ renders the luminosity constraints irrelevant for relatively large couplings, say $g \gtrsim 10^{-5} - 10^{-4}$. When the scalar is heavier, the scattering cross section tends to be smaller and thus the coupling $g$ has to be larger for a fixed value of $D_S$. The supernova core contains more $\nu_e$ than the muon and tauon flavours to scatter with the scalar, thus the trapping effect is stronger for the electron flavor and the MFP condition $D_S < R_c$ excludes larger region of the luminosity constraint for the couplings involving the electron flavour.


Combining the analysis of luminosity and MFP, if the couplings $g$ are too large, after produced, the scalars get easily trapped in the inner core by scattering with the neutrino bath and thus can not transport any energy outside the core and lead to extra energy loss. However, as depicted in Fig.~\ref{fig:lum}, a large range of couplings still remains excluded from the simple examination of energy loss: $10^{-5} \lesssim g_{\alpha\beta} \lesssim 10^{-11}$, depending on the neutrino flavours involved and the scalar mass.

\section{Deleptonization constraints}
\label{sec:delep}


The scalar production in the supernova core can not only lead to exotic energy loss but also deleptonize the core in all the three flavours. In particular,
supernova simulations reveal that the electron lepton number $Y_{L_e}$ is severely constrained~\cite{Kachelriess:2000qc}: A successful explosion requires that at the time of core bounce $Y_{L_e} \gtrsim 0.375$ and only a small variation of $0.015$ of $Y_{L_e}$ is allowed~\cite{Bruenn:1985,Baron:1987zz,Kachelriess:2000qc}. For SN1987A, the total energy loss due to neutrino emission is about $10^{53}$ erg, and the average neutrino energy is about 10 MeV, thus it is expected that a total number of $10^{57}$ of $\nu_e$ was emitted from SN1987A within a duration of a few seconds. The electron lepton number can thus be used to constrain the electron flavour relevant couplings $g_{e\beta}$ as a function of the scalar mass.


To estimate the deleptonization effect due to presence of the scalar, we count how many {scalars} are produced and then decay outside the supernova core. The formula for the number production rate is rather analogous to Eq.~(\ref{eqn:production}), i.e.
\begin{eqnarray}
N_{S} &=& \int d\Pi_3 \ F_S \sum_{\rm spins}|\mathcal{M}|^2(2\pi)^4 \delta^{(4)}(p_1+p_2-p_S)f_1f_2 \,.
\end{eqnarray}
The only difference lies in the absence of the energy factor $E_S$ (Note that the decay factor has also to be included to exclude those scalars decaying inside the core). The total number of scalars emitted is then $\mathcal{N}_S = V_{\rm core} N_S \Delta t$, with $\Delta t$ the duration of neutrino burst. The deleptonization constraints are presented in Fig.~\ref{fig:lum} (right panel), where the constraints apply only to the entries of $g_{e\beta}$ relevant to the electron flavor. In the calculation of Fig.~\ref{fig:lum} we have assumed $\Delta t = 1$ sec; a larger $\Delta t$ {implies a smaller lepton number variation} and pushes the constraints on $m_S$ and $g_{\alpha\beta}$  {to be} more stringent.

\section{Dimension-5 interactions}
\label{sec:dim5}

In this section we explore the possibility that the  interaction of the (pseudo)scalar with neutrinos, instead of being at the renormalizable level,  is encoded by effective dimension-5 operators, which can be seen as the low energy limit of some UV complete theories. The operator we consider here is taken to be of the form
\begin{eqnarray}
\label{eqn:coupling5}
\mathcal{L} = \frac{h_{\alpha\beta}}{4 \Lambda_{\rm TeV}} (\nu^T_\alpha i\sigma_2 \nu_\beta) (SS) + {\rm h.c.} \,,
\end{eqnarray}
where $h_{\alpha\beta}$ are the flavor dependent couplings and $\Lambda_{\rm TeV} = 1$ TeV is the universal cutoff scale explicitly set for concreteness\footnote{Note that a different cut-off scale can be easily adapted by a simple rescaling of the coupling $h$.}. In this case, in contrast to the renormalizable couplings in Eq.~(\ref{eqn:coupling4}), the scalar $S$ can only be pair produced via neutrino annihilation $\nu_\alpha \nu_\beta \to SS$.
In the absence of ``trilinear'' couplings in Eq.~(\ref{eqn:coupling4}), the decay of $S$ is forbidden, and the scalars can only annihilate back into neutrinos or scatter on the neutrino bath.

Since only a small portions of neutrinos are expected to annihilate into scalars inside the supernova core, the number density of $S$ is negligible compared to the one of neutrinos, thus in calculation of the emission rate {we} safely neglect the inverse process $SS \to \nu\nu$. To calculate the energy emission rate, and then derive  {analogous} luminosity constraints as in Section~\ref{sec:lum}, we make the replacement in Eq.~(\ref{eqn:production}):
\begin{eqnarray}
d \Pi_3 \to d \Pi_4 \,, \quad
E_S \to E_{S_1} + E_{S_2} \,.
\end{eqnarray}
with the $\delta$-function and the amplitude altered accordingly.
After integrating over the phase space, we obtain the emission rate for the dimension-5 operator:
\begin{eqnarray}
\label{eqn:Q5}
Q_{}
&=& \frac{F_S |h_{\alpha\beta}|^2 T^7}{2^7 \pi^5 \Lambda^2_{\rm TeV}}
\int_{0}^{\infty} du \int_{0}^{\infty} dv
\int_{-1}^{1} dz_i \int_{-1}^{1} dz_f  \,
\sqrt{uv} w \nonumber \\
&& \times \frac{(I_1+I_2)I_5}{I_1I_2I_6
\left( 1 + e^{I_1 - x_1} \right)
\left( 1 + e^{I_2 - x_2} \right)} \,
\Theta(I_1 + I_2 - 2 \sqrt{q+u}) \,.
\end{eqnarray}
The $\Theta$-function ensures that the {total energy} carried by the incoming neutrinos are large enough to produce the two massive scalars. The dimensionless parameters $u$, $v$, $w$, $q$, $z_{i,\,f}$, $x_i$ and the functions $I_{i}$ are defined in the Appendix.



\begin{figure*}[t]
  \centering
  \includegraphics[width=0.48\textwidth]{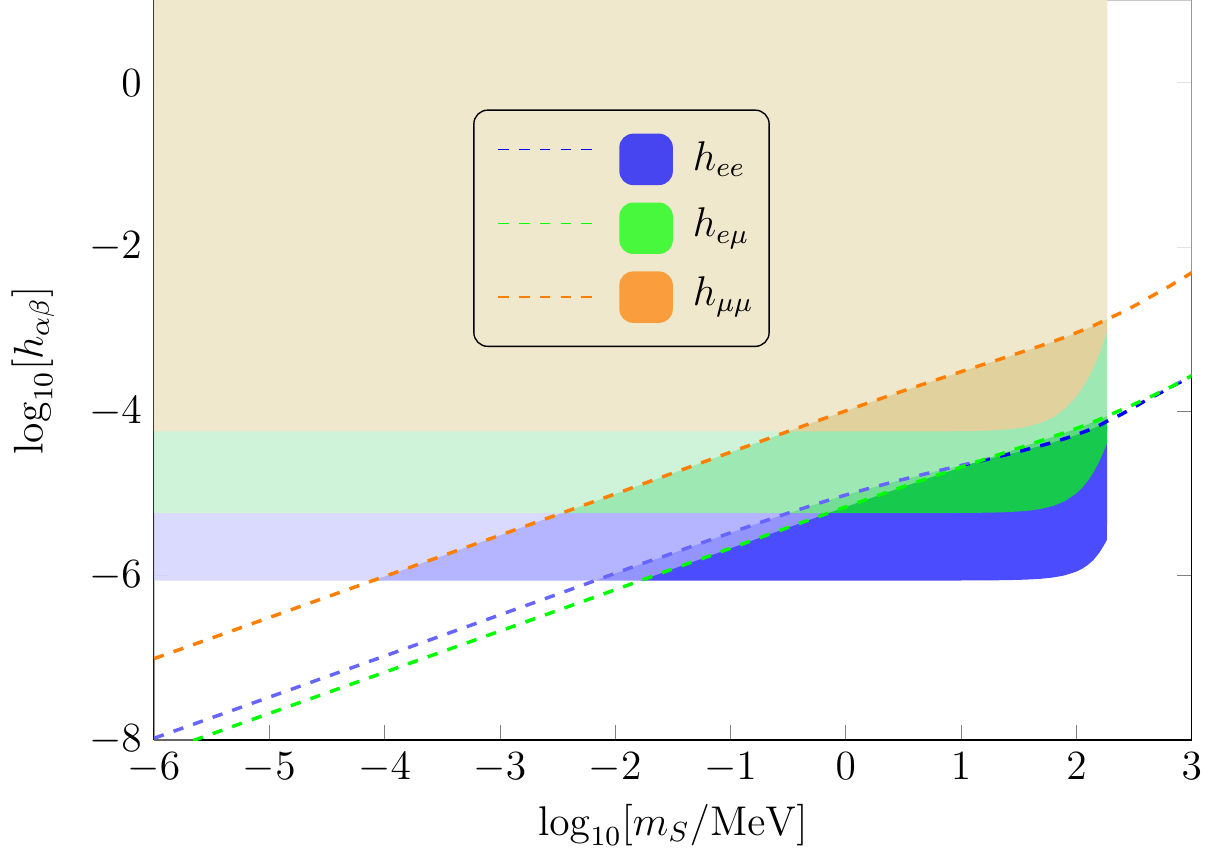}
  \includegraphics[width=0.48\textwidth]{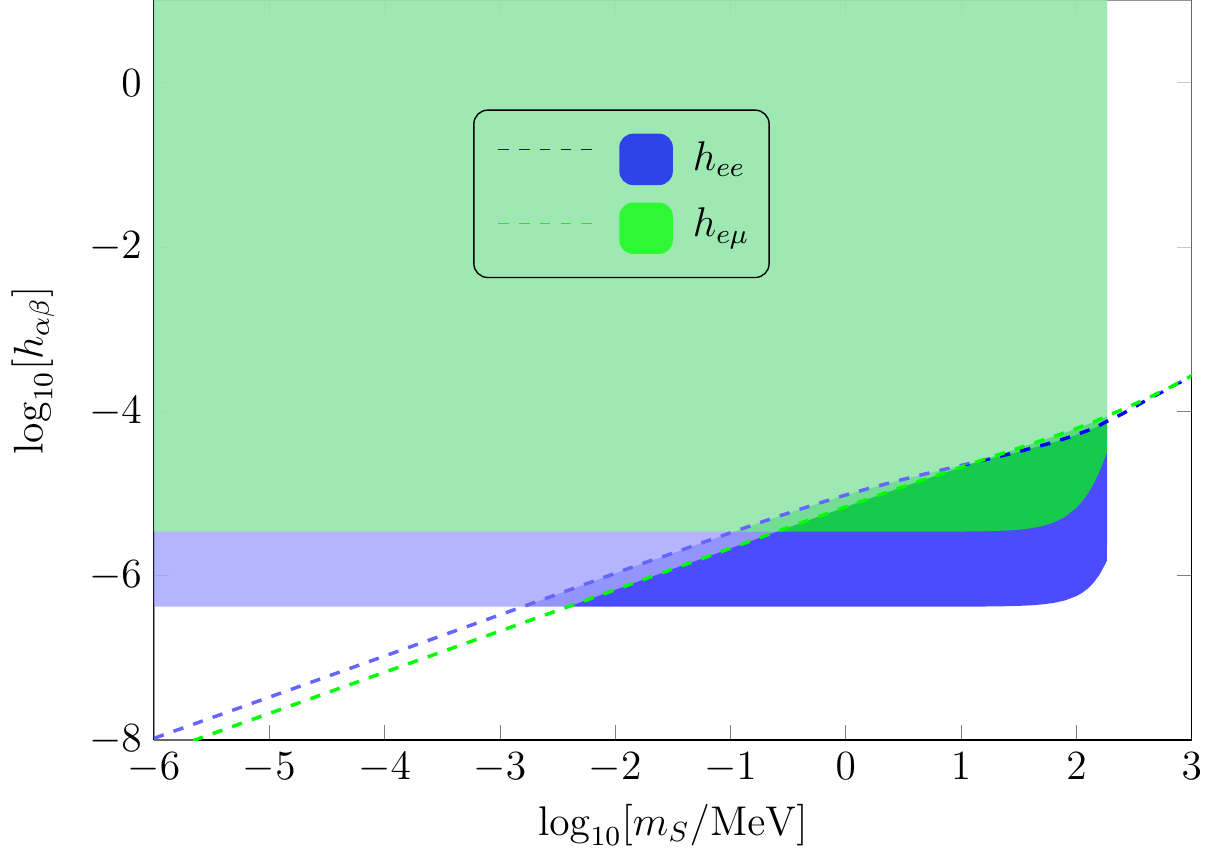}
  \caption{\label{fig:lum3}
  \footnotesize
  Similar constraints as in Fig. \ref{fig:lum}, for the dimension-5 operator in Eq.~(\ref{eqn:coupling5}).
  }
\end{figure*}

Since the scalar $S$ does not decay and the annihilation into neutrinos {is} neglected, all the {scalar particles} produced in the core are taken into account, as long as the mean free path is longer than the core size. The excluded regions from luminosity and deleptonization with regard to the different flavor combinations are shown respectively in the left and right panel of Fig.~\ref{fig:lum3}.

As far as the {trapping effect} is concerned, it is still driven by the process $\nu_\alpha S \to \nu_\beta S$, which in this case is a contact interaction. The calculation of the MFP in {Section \ref{sec:mfp}} thus holds true when using the appropriate scattering cross section and the $f(E_S)$ in Eq.~(\ref{eqn:fES}) is halved. The regions  { where}  $D_S < R_C$ are {surrounded from below by the dashed lines in Fig.~\ref{fig:lum3}, and indicate where the luminosity and deleptonization constraints are not applicable}.
When the trapping effect is considered, it is found that only small regions could be excluded from both the energy loss and deleptonization arguments, with the scalar mass ranging from MeV to 200 MeV, and the coupling $h$ going from $10^{-6}$ to $10^{-2}$.

\section{Prospects for future experiments}
\label{sec:prospects}

The main weakness of the constraints from SN1987A lies in the very poor experimental data, from which the only possible constraints one can set are mostly   rough estimations of orders of magnitude.
If a supernova explosion could happen nearby in the next decades, the detection of a supernova burst would be incredibly more efficient. Indeed, detections of neutrinos by Super-Kamiokande and IceCube, as well as the next generation dark matter direct detection experiments \cite{Abbasi:2011ss,Lang:2016zhv} could allow to detect thousands to millions of events, depending on how far the supernova is.
In this section we estimate to what extent such improvements of detections could reinforce our constraints.

In this paper we make the assumption that the emission of exotic scalar particles would not alter too much the thermal distribution and explosion dynamics of the neutrino bath during the whole period of supernovae evolution including the phase of deleptonization. Such assumption is  very strong
and should be overcome by running simulations of supernova explosion including these additional interactions of neutrinos.
We postpone such a precise study for future works and propose  {simply} here a  {rough} estimation of    {the expected} constraints as a motivation for the appropriate simulations.


If we write $N_{\rm event}$ the total number of neutrinos detected during a time bin of ${\Delta t}$, and assuming an error bar of order $\sqrt{N_{\rm event}}$, then the constraint on the luminosity of massive scalars within $\Delta t$  {can be written}
\begin{equation}\label{eq:Ncount}
\mathcal L_{S}^{\Delta t}\ <\ \frac{1}{\sqrt{N_{\rm event}}}\ \mathcal{L}_{\nu,\rm ~simu}^{\Delta t}\;,
\end{equation}
where $\mathcal{L}_{\nu, \rm ~simu}^{\Delta t}$ is the flux of neutrinos integrated over ${\Delta t}$ predicted by simulations.

\begin{figure}[t]
  \centering
  \includegraphics[width=0.5\textwidth]{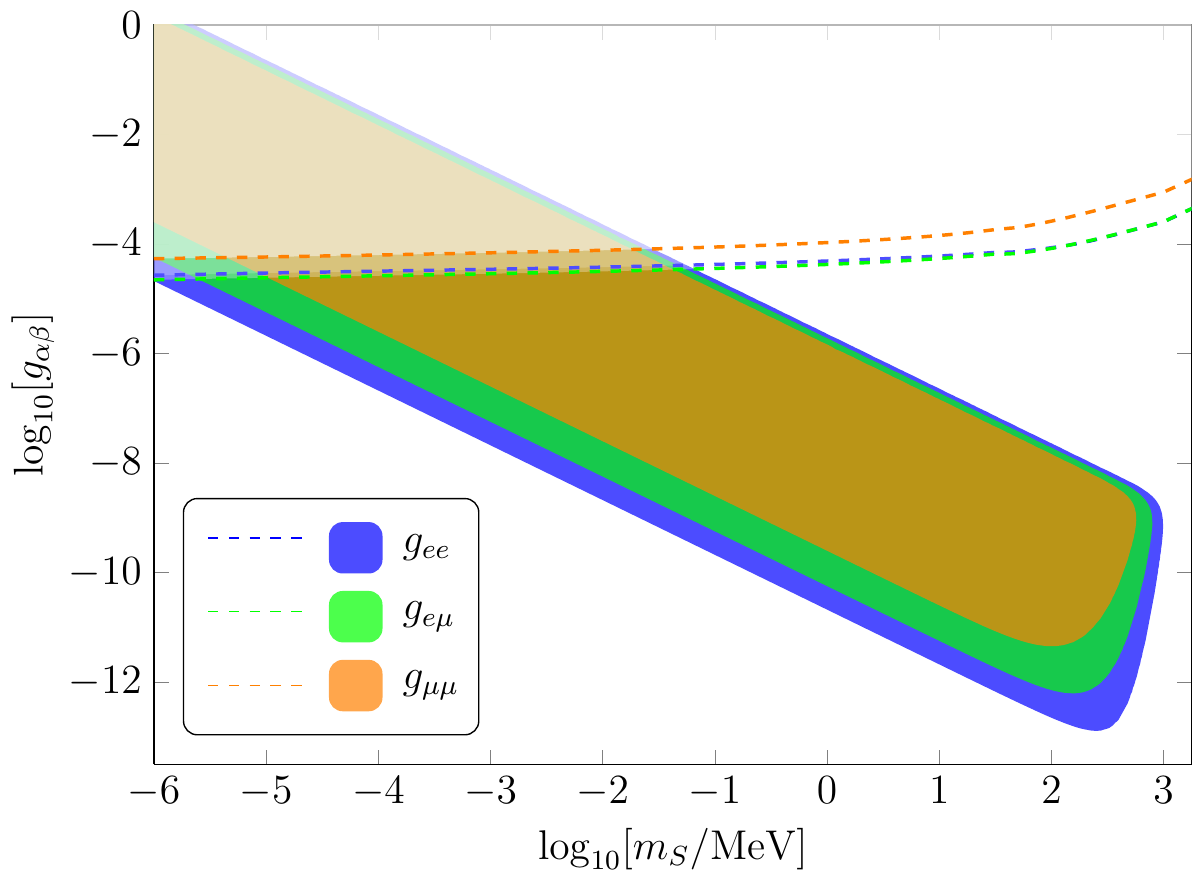} \\
  \caption{\label{fig:lum2}
  \footnotesize
  Neutrino luminosity constraints on the coupling $g_{\alpha\beta}$, as functions of the scalar mass $m_S$, assuming the detection of a nearby supernova ($d \simeq 0.2 \, \mathrm{kpc}$) by IceCube, with a number of $10^8$ of events collected within 0.1 second.}
\end{figure}

\begin{figure}[t]
  \centering
  \includegraphics[width=0.5\textwidth]{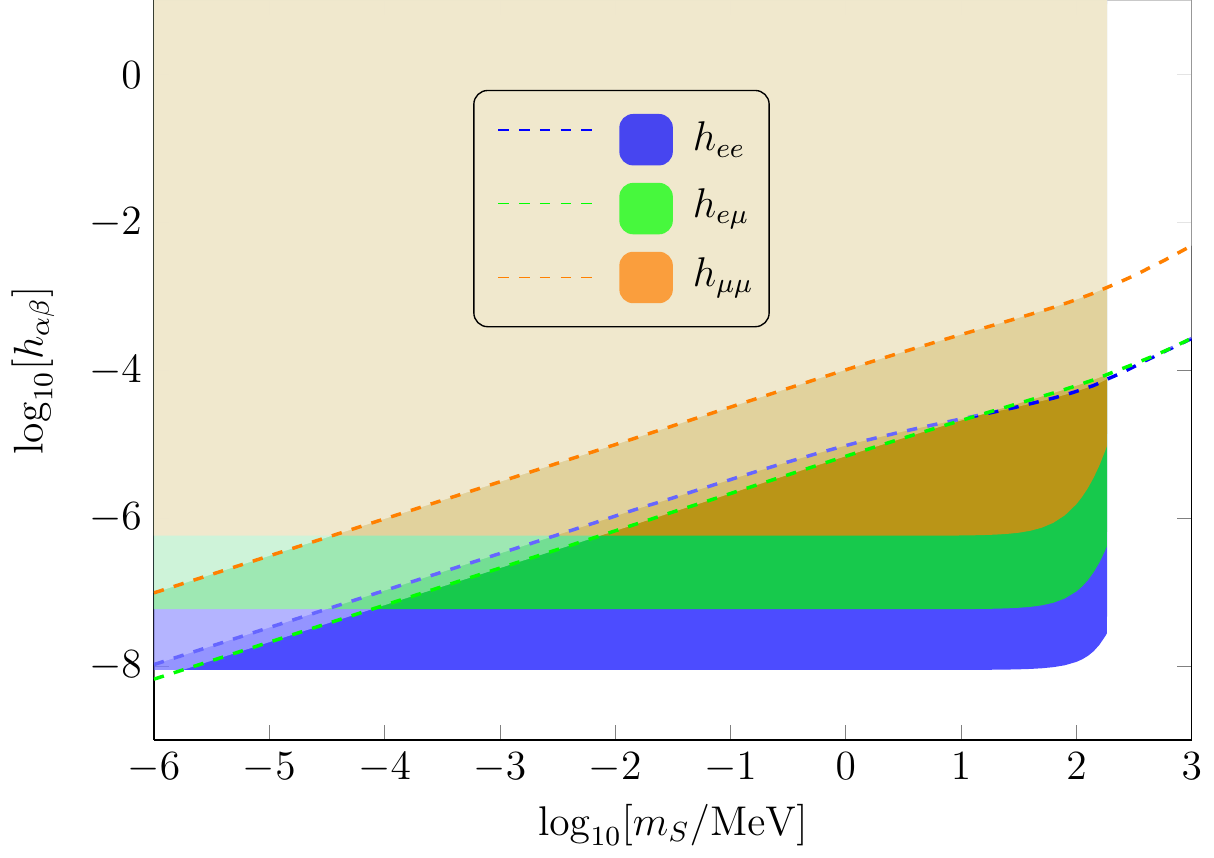} \\
  \caption{\label{fig:prospects2}
  \footnotesize
  Similar constraints as in Fig.~\ref{fig:lum2}, for the dimension-5 operator in Eq.~(\ref{eqn:coupling5}).}
\end{figure}

In what follows we use the reference simulations
of~\cite{Fischer:2016cyd} in the case of an $18 M_{\odot}$ supernova. We focus in particular on the most optimistic case of a nearby supernova located at $d \simeq 0.2 \, \mathrm{kpc}$. In such a case the luminosity flux reaches $\sim 10^{53} \, \mathrm{erg / sec}$ during approximately 0.1 sec after the core bounce.
Within the time bin of 0.1 sec, one could conservatively expect a total number of $10^5$ events in the Super-Kamiokande detector and up to $10^8$ at IceCube.
One can hence impose the {constraints }
\begin{equation}
\mathcal L_{S}\ \lesssim\
\begin{cases}
3\times 10^{50}\ {\rm erg/sec} & \text{for Super-K} \\
5\times 10^{48}\ {\rm erg/sec} & \text{for IceCube}
\end{cases}
\end{equation}
Using these constraints, we can reinforce the limits in Section~\ref{sec:lum} and \ref{sec:delep}.
The IceCube prospects for the dimension-4 and dimension-5 couplings are depicted respectively in Fig. \ref{fig:lum2} and \ref{fig:prospects2}. As expected, more data collection could set more stringent limits and exclude larger regions in the parameter space. For the dimension-4 coupling, the coupling $g_{\alpha\beta}$ can be probed even down to $10^{-13}$, while for the dimension-5 case, the probable scalar mass range spans 6 orders of magnitude with the coupling $h_{\alpha\beta}$ spanning almost 4 orders of magnitude. With less data collected at Super-Kamiokande, the limits go less stringent, but still exclude much large regions as for the IceCube limits.

\section{Conclusion}
\label{sec:conclusion}

Core-collapse supernovae provide a unique circumstance to study the beyond SM particles and couplings, and have been  {extensively studied} with regard to neutrinos, axion, Majoron etc. In this letter we demonstrate to what extent the couplings of a massive scalar (or pseudoscalar) to SM neutrinos can be constrained from the supernova side. This is well motivated from the large variety of Majoron models, and applies to the most general cases. Two distinct types of couplings are considered, which are respectively dimension-4 and dimension-5, as shown in Eqs.~(\ref{eqn:coupling4}) and (\ref{eqn:coupling5}). We apply two different constraints on both {the} couplings: the first one concerns the energy loss of supernovae due to the (pseudo)scalar emission and the second one is the electron lepton number depletion in the supernova core. The limits from SN1987A on the dimension-4 couplings are collected in Fig.~\ref{fig:lum}, with the relatively dark coloured regions excluded in the parameter space of (pseudo)scalar mass and couplings. The possibility of scattering with neutrinos inside the core, which tends to trap the scalar particles, is also taken into account. The exclusion regions can be summarized as follows
\begin{align*}
2.1\times 10^{-9} \, {\rm MeV}
&\lesssim  {|g_{ee}|}\times{m_S}
\lesssim 1.6\times 10^{-6} \, {\rm MeV} \,,\\
5.5\times 10^{-9} \, {\rm MeV} &\lesssim {|g_{\mu\mu}|}\times{m_S}\lesssim 1.1\times 10^{-6} \, {\rm MeV}\,,\\
2.3\times 10^{-8} \, {\rm MeV} &\lesssim {|g_{e\mu}|}\times{m_S}\lesssim 6.6\times 10^{-7} \, {\rm MeV} \,,
\end{align*}
in the region where, roughly, $m_S \in [10,\, 500] \, \mathrm{MeV}$. The deleptonization constraints overlap largely with the corresponding limits from energy loss. In the case of dimension-5 operator, we can constrain the parameter space for a scalar mass from MeV to 100 MeV and couplings $h_{\alpha\beta}$ from $10^{-6}$ to $10^{-2}$, depending on the neutrino flavours and scalar mass. The deleptonization constraints give again similar bounds for coupling involving the electron flavour.

For what concerns future observability, we derive also the prospects in the case of a nearby supernova explosion. Given a huge amount of neutrino data collected in future experiments like {Super-Kamiokande} and IceCube, the current limits {can be largely} improved, as shown in Fig.~\ref{fig:lum2} and \ref{fig:prospects2}.
A non-deviation {of experimental data} with respect to {supernova} simulations could  {exclude} {large} regions of the parameter space.
Probing models standing in this region of parameter space may hence be rendered possible by future supernova observations.

Our estimations of supernova constraints on the couplings of massive (pseudo)scalars to neutrinos are complementary to terrestrial experiments such as those from meson and lepton decay, which are both,  {in some sense,} the counterpart of constraints on couplings of a massless (pseudo)scalar to the SM neutrinos.
Finally, more {involved} simulations of supernova explosions, including the emission of massive scalars, would be of great interest for constraining such secret interactions, as done for the case of axions in~\cite{Fischer:2016cyd}.




\section*{Acknowledgement}
The authors would like to thank D. Teresi for valuable discussions and collaboration in the early stage of this work, and also T. Hambye for the suggestive remarks. L.H. would like to thank A. Payez and  S. Chakraborty for useful discussions and reference suggestions. Y.Z. is grateful to R. N. Mohapatra for the enlightening discussions on the Majoron models. The work of L.H. and Y.Z. is funded by the Belgian Federal Science Policy through the Interuniversity Attraction Pole P7/37. Y.Z. is also grateful to Hong-Hao Zhang for his gracious hospitality during the visit at Sun Yat-Sen University where part of the work was done and it is supported by the National Natural Science Foundation of China (NSFC) under Grant No. 11375277.

\appendix
\section{Energy emission rate for the dimension-5 operator}
\label{sec:app}

The calculation procedure of the emission rate for the dimension-5 operator is very similar to that in ref~\cite{Kazanas:2014mca}, with the dimensionless parameters defined as:
\begin{eqnarray}
&& u \equiv \frac{{\bf P}^2}{T^2} \,, \quad
   v \equiv \frac{{\bf p}_i^2}{T^2} \,, \quad
   w \equiv \frac{{\bf p}_f^2}{T^2} \,, \nonumber \\
&& q \equiv \frac{m_S^2}{T^2} \,, \quad
   x_i \equiv \frac{\mu_i}{T} \,, \quad
   z_{i,f} \equiv \cos\theta_{i,f} \,,
\end{eqnarray}
with ${\bf P}$, ${\bf p}_i$ and ${\bf p}_f$ the momenta defined in the center-of-mass frame of the annihilating neutrinos which are related to the momenta of neutrinos (${\bf p}_{1,2}$) and scalars (${\bf p}_{S_1, S_2}$) via
\begin{eqnarray}
&& {\bf p}_1 = {\bf P} + {\bf p}_i \,, \quad
   {\bf p}_2 = {\bf P} - {\bf p}_i \,, \nonumber \\
&& {\bf p}_{S_1} = {\bf P} + {\bf p}_f \,, \quad
   {\bf p}_{S_2} = {\bf P} - {\bf p}_f \,.
\end{eqnarray}
$\theta_{i(f)}$ is the angles between ${\bf P}$ and ${\bf p}_i$ (${\bf p}_f$), and the $I_i$ functions are defined as follows:
\begin{eqnarray}
I_1 &\equiv& \frac{E_1}{T} = \sqrt{u+v+2\sqrt{uv}z_i} \,, \\
I_2 &\equiv& \frac{E_2}{T} = \sqrt{u+v-2\sqrt{uv}z_i} \,, \\
I_3 &\equiv& \frac{E_{S_1}}{T} = \sqrt{q+u+w+2\sqrt{uw}z_f} \,, \\
I_4 &\equiv& \frac{E_{S_2}}{T} = \sqrt{q+u+w-2\sqrt{uw}z_f} \,, \\
I_5 &\equiv& \frac{E_1 E_2 - {\bf P}^2 + {\bf p}_i^2}{T^2}\,, \nonumber \\
&=& \sqrt{(u-v)^2 + 4uv(1-z^2)} - (u-v) \,, \\
I_6 &\equiv& |\sqrt{w} (I_1 + I_2) + \sqrt{{u}} z_f (I_4-I_3)| \,.
\end{eqnarray}
the parameter $w$ are related to other quantities via the equation
\begin{eqnarray}
w = \frac{(I_1+I_2)^2 [(I_1+I_2)^2 - 4(q+u)]}{4 [(I_1+I_2)^2 - 4uz_f^2]} \,.
\end{eqnarray}

\end{document}